\documentclass[%
 reprint,
 amsmath,amssymb,
 aps,nofootinbib,
  twocolumn,
]{revtex4-1}
\usepackage[usenames, dvipsnames]{color}
\usepackage[autostyle]{csquotes}
\usepackage{pgfplots}
\usepackage{tikz}
\usepackage{graphicx}
\usepackage{dcolumn}
\usepackage{bm}
\usepackage{enumerate}

\begin{document}

\preprint{APS/123-QED}

\title{End-to-End AI-Based Point-of-Care Diagnosis System for Classifying Respiratory Illnesses and Early Detection of COVID-19}

\author{Abdelkader Nasreddine Belkacem$^{1}$}
\email{Correspondence: belkacem@uaeu.ac.ae}
\author{Sofia Ouhbi$^{2}$}%
\author{ Abderrahmane Lakas$^{1}$}%
\author{ Elhadj Benkhelifa$^{3}$}%
\author{ Chao Chen$^{4}$}%

\affiliation{$^{1}$%
 Department of Computer and Network Engineering, CIT, UAEU, UAE
}%
\affiliation{$^{2}$%
 Department of Computer Science and Software Engineering, CIT, UAEU, UAE
}%

\affiliation{$^{3}$%
Cloud Computing and Applications Reseach Lab, Staffordshire University, UK
}%

\affiliation{$^{4}$%
 Key Laboratory of Complex System Control Theory and Application, Tianjin University of Technology, China
}%

\date{\today}

\begin{abstract}
 Respiratory symptoms can be a caused by different underlying conditions, and are often caused by viral infections, such as Influenza-like illnesses or other emerging viruses like the Coronavirus. These respiratory viruses, often, have common symptoms, including coughing, high temperature, congested nose, and difficulty breathing. However, early diagnosis of the type of the virus, can be crucial, especially in cases such as the recent COVID-19 pandemic. One of the factors that contributed to the spread of the pandemic, was the late diagnosis or confusing it with regular flu-like symptoms. Science has proved that one of the possible differentiators of the underlying causes of these different respiratory diseases is coughing, which comes in different types and forms. Therefore, a reliable lab-free tool for early and more accurate diagnosis that can differentiate between different respiratory diseases is very much needed. This paper proposes an end-to-end portable system that can record data from patients with symptom, including coughs (voluntary or involuntary) and translate them into health data for diagnosis, and with the aid of machine learning, classify them into different respiratory illnesses, including COVID-19. With the ongoing efforts to stop the spread of the COVID-19 disease everywhere today, and against similar diseases in the future, our proposed low cost and user-friendly solution can play an important part in the early diagnosis.
 
\begin{description}
\item[Keywords]
COVID-19; Intelligent learning; Respiratory illness; Health diagnosis, e-health 
\end{description}
\end{abstract}

\maketitle


\section{Introduction}

People usually take breathing and respiratory health for granted and often forget that their lungs are vital organs, which are vulnerable to infections and injury. According to the World Health Organization (WHO), respiratory diseases are among the leading causes of death and disability in the world \cite{firs2017}. Respiratory conditions include “acute respiratory infections as well as chronic respiratory diseases, such as asthma, chronic obstructive pulmonary disease and lung cancer” \cite{emro}.  This can be aggravated by multiple determinants such as: direct and indirect exposure to tobacco smoke; heavy exposure to air pollution; occupational related disorders; malnutrition and low birth weight, but most commonly by exposure to virus such as influenza virus or Coronaviruse \cite{smith1993fuel}. Making a timely and correct diagnosis is essential for treatment as symptoms of respiratory illnesses are often very similar to each other \cite{badnjevic2013integrated}, which can cause confusions that can lead to misdiagnosis. This can result in some cases, such as the the recent COVID-19 pandemic, in catastrophic consequences of further spread of the infection. Therefore, ensuring a diagnostic differentiator is highly crucial for timely and accurate prognosis and appropriate actions \cite{tang2020laboratory}.

Cough is considered a key symptom of respiratory diseases \cite{cho2016respiratory}. Cough is a natural respiratory defence mechanism to protect the respiratory tract and one of the most common symptoms of pulmonary disease \cite{korpavs19792}. A cough is normally initiated with an inspiration of a variable volume of air, followed by closure of the glottis, and contraction of the expiratory muscles that compresses the gas in the lungs \cite{abaza2009classification}. These events occur immediately before the sudden reopening of the glottis and rapid expulsion of air from the lungs. When flow limitation is reached during coughs that begin at the same lung volume, the airflow and acoustic properties are repeatable and unique for a given subject \cite{day2004identification}. There are two types of cough: productive and non-productive \cite{murata1998discrimination}. A productive cough produces phlegm or mucus, clearing it from the lungs, while a non-productive cough, also known as a dry cough, does not produce phlegm or mucus. Analyzing the cough sound can give information about the pathophysiological mechanisms of coughing by indicating the structural nature of the tissue during therapy that leads to certain patterns of cough \cite{korpavs1996analysis}. Similarly, the character of the cough sound gives information about the behaviour of the glottis and whether the glottis behaves differently in different pathological conditions \cite{korpavs1996analysis}. Analysis of the cough sound record has significant value in prognosis because its changes may indicate the effectiveness of therapy or the progress of disease \cite{korpavs1996analysis}. 

There is a growing interest in using the characteristics of voluntary and involuntary (i.e. spontaneous or reflex) coughs to detect and characterize lung disease \cite{everett2007chronic,everett2007chronic}. Automated real-time and reliable lab-free tools for cough characterisation and classification could be valuable for timely and accurate diagnosis and differentiating between different respiratory illnesses, which is crucial for timely and correct treatment \cite{larson2011accurate}. This can be particularly useful in parts of the world with limited access to laboratory resources \cite{world1990acute}. Since coughs are often seasonal events, a cough detector/classifier has to have a very low false alarm rate in order to be clinically reliable. Such a system needs a high sensitivity in order to detect the infrequent event of a cough \cite{amoh2015deepcough}. To the best of our knowledge, there is no standard method for automatically evaluating coughs that has been established, even though a few approaches have been reported in literature \cite{korpavs1996analysis,korpas2003analysis}. We also recognise that to achieve accurate diagnosis, data of other accompanying symptoms, such as temperature, must be used in conjunction with the cough data. 

This paper proposes a portable end-to-end point-of-care system, supported by Artificial Intelligence (AI) module for classifying and diagnosing different respiratory illnesses, including early detection of COVID-19. The novel proposed system is composed of hardware and software components. The system will be able to record patients' or users' symptoms, including body temperature, cough sound, and airflow, using specialist sensors. The recorded data will then be translated to health data, which will be processed by the machine learning module, to find patters and classify the combined symptoms for different respiratory conditions, including COVID-19. A customised mobile application (app) will be developed and used for data processing and visualisation. The app will allow the users to interact with the system's parameters, including the option to submit their results to the doctors electronically. Patients' data will be stored securely in the cloud. Figure 1 provides a high level illustration of the proposed system.  

The remainder of this paper is as follows: Section \ref{related-work} presents related works identified in the literature. In Section \ref{sec: architectire}, we explain the main components of the proposed system and its detailed architecture. In Section \ref{Conclusions}, we conclude by discussing the prospective of the proposed framework, the challenges and the limitations.

\begin{figure*}[]
\begin{center}
\includegraphics[width=15cm]{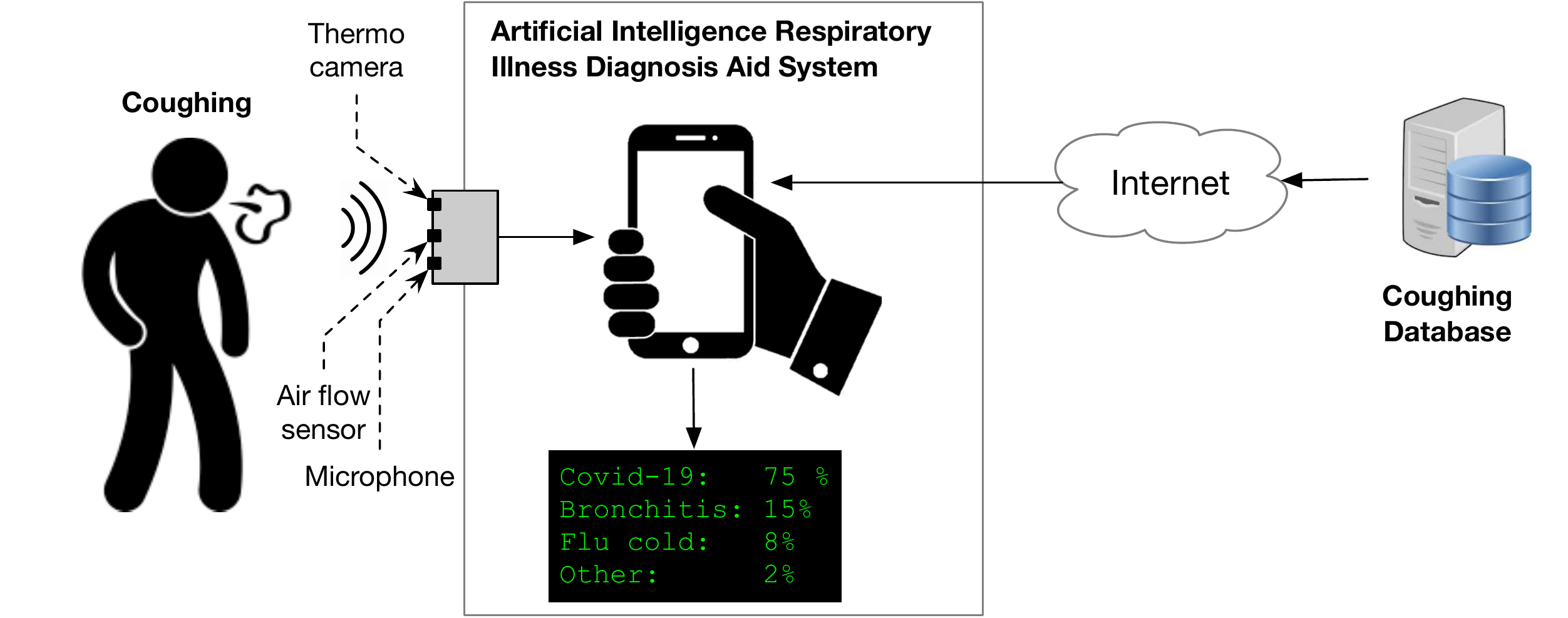}
\caption{Illustration of the proposed AI-based diagnosis aid system for simultaneously recording and classifying cough characteristics}
\end{center}
\end{figure*} 
\section{Related Work}
\label{related-work}

Several studies have been conducted to classify and detect lung-related diseases using AI. Liu et al. \cite{liu2017lung} proposed a lung sound classification algorithm based on Hilbert-Huang transform features and multilayer perceptron network for noninvasive diagnosis of pulmonary diseases. The algorithm was tested using the R.A.L.E. database with a multi-layer perceptron classifier and achieved an averaged classification accuracy of 95.84\%. Aykanat et al. \cite{aykanat2017classification} proposed a non-invasive method of classifying recorded respiratory sounds by an electronic stethoscope. They recorded with this device 17,930 lung sounds from 1,630 subjects. Their results showed that using convolutional neural network (CNN) and support vector machine (SVM) machine learning algorithms accurately classified and pre-diagnose respiratory audio. Azam et al. \cite{azam2018smartphone} presented a scheme to detect irregular patterns occurred in respiratory cycles due to respiratory diseases. They used 255 breath cycles captured using a smartphone under natural setting. Their experiments showed an accuracy around 75\% using SVM for asthmatic inspiratory cycles and complete respiratory sounds.

Few AI systems have been developed to detect respiratory illnesses. Among them, \textit{FluSense} \cite{al2020flusense}, which is a contactless syndromic surveillance platform for influenza-like illness used in hospital waiting areas. The aim of \textit{FluSense} is to expand the current paradigm of influenza-like illness surveillance by capturing crowd-level bio-clinical signals directly related to physical symptoms of influenza-like illness from hospital waiting areas in an unobtrusive and privacy-sensitive manner. \textit{FluSense} uses a microphone array and a thermal camera along with a neural computing engine to passively and continuously characterize speech and cough sounds along with changes in crowd density on the edge in a real-time manner. The researchers conducted an IRB-approved 7 month-long study from December 10, 2018 to July 12, 2019, where they deployed \textit{FluSense} in four public waiting areas within the hospital of a large university \cite{al2020flusense}. During this period, the \textit{FluSense} platform collected and analyzed more than 350,000 waiting room thermal images and 21 million non-speech audio samples from the hospital waiting areas. The study \cite{al2020flusense} showed that \textit{FluSense} accurately predicted daily patient counts with a Pearson correlation coefficient of 0.95. The \textit{FluSense} platform did not take into consideration all respiratory illnesses nor additional health data. Cough data is important and relevant features but not sufficient one to be used for all respiratory illnesses.   



With the recent rise in the new Coronavirus pandemic, early and accurate testing has been crucial due to the fast spreading feature of this virus, which caused close to 10 million cases and almost 500,000 deaths up to late June 2020 \cite{covid-june}. Among the important factors for COVID-19 spread have been lack of testing and erroneous diagnosis, which might be due to inaccurate testing or confusion with flu-like symptoms. These challenges were recognised at early days of the pandemic \cite{wu2020characteristics}. Two main mechanisms to detect the Coronavirus disease were adopted  \cite{kang2020recent}: (i) clinical analysis of chest computed tomography (CT) scan images, and (ii) blood test results. The confirmed COVID-19 patients, commonly, manifest persistent fever, tiredness, and dry cough. Although little literature exists on the diagnosis of COVID-19 given its recent emergence, there have been a rapid response in the research community toward COVID-19 diagnoses and prediction using AI-based software on medical imaging \cite{shi2020review,wynants2020prediction}.
Chen et al. \cite{chen2020deep} developed an AI engine based on deep learning to detect COVID-19 disease using high resolution CT images. It has been deduced in \cite{wynants2020prediction} that the prediction models for COVID-19 diagnosis are poorly reported, and at high risk of bias. Therefore, there is a need to combine these models with other diagnostic methods such as lab tests. However, these methods are costly and time consuming, which was problematic for many countries which do not have the capacity to accommodate tests for larger populations. Cheaper rapid tests have emerged recently to increase the testing capacity of many countries \cite{al2020testing}. However, these low-cost and rapid tests are often non-trusted, with many cases of false positives or false negatives. For this reason, several researchers have started looking into alternative solutions to detect COVID-19. Maghdid et al. \cite{maghdid2020novel} have proposed a novel framework on how to detect COVID-19 using on-board smartphone sensors. However, their idea is still in the conceptualization phase and has not yet been implemented. The proposed solution is designed for certain types of smartphones, which may not be available and affordable to a large number of people. In addition their proposed framework is designed for COVID-19 diagnosis only.  

It is therefore important to develop an End-to-End point-of-care diagnosis devise, that can collect data of combined symptoms common in respiratory illnesses (eg. cough, body temperature, and airflow) and classify them as a diagnostic differentiator, detecting for instance COVID-19. The system should also be able to be adaptable to newly emerging illnesses for early testing.  




\section{The Proposed System}
\label{sec: architectire}

The objective of cough classification is to develop an automatic system that is capable of classifying various attributes of coughs such as the intensity of the cough, time-frequency energy distributed, or whether the cough is wet or dry. Different respiratory diseases such as Bronchitis, Tuberculosis, Asthma, etc. may have different effects on the pulmonary system and are, therefore, identifiable through the changes observed on the sounds of coughs. For instance, coughs from asthmatic patients tended to have different energy signatures than that from non-asthmatic patients. In particular, asthmatic coughs exhibit more energy in the low frequency \cite{al2013signal}. On the other hand, the study of the phases of the cough reveal that dry coughs have lower energy compared to wet coughs in phase 2. In addition, during this phase, most of the signal energy is contained between 0-750 Hz for the wet coughs, and 1500-2250 Hz for dry coughs \cite{pramono2016cough}. Thus, most of cough recording experiments have been using a sampling frequency of 22,050 Hz to cover all cough types. Before designing the architecture of proposed system, it is important, first, to understand the cough audio preprocessing phases, which are described below.

\begin{figure}[]
    \centering
    \includegraphics[width=0.9\linewidth]{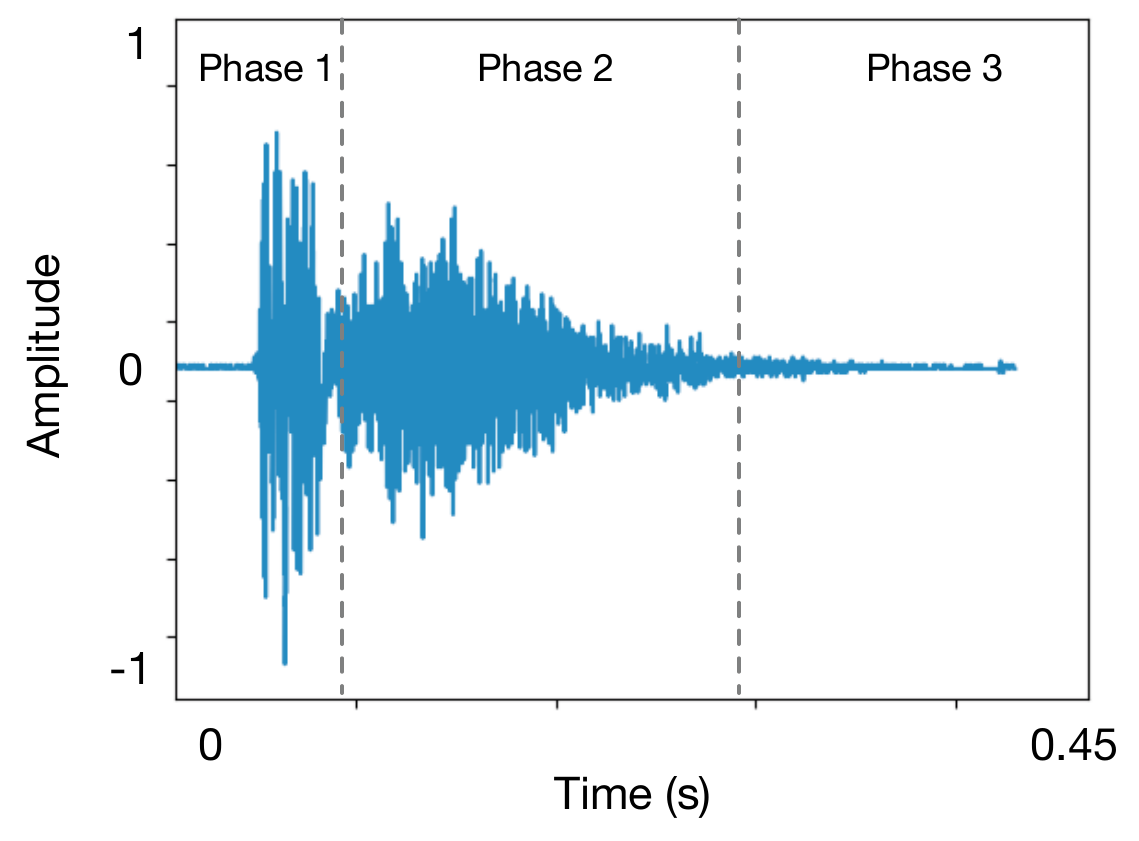}
    \caption{A typical cough sound structure. There are three common patterns of cough based on the number of phase such as three-phase cough, two-phase cough, and peal cough.}
    
    \label{coughstructure}
    
    \vspace{-4mm}
\end{figure}
\vspace{-4mm}
\subsection{Cough Audio Preprocessing}

\subsubsection{Cough audio analysis}
The acoustic sound of a cough is generated by the contractions of the respiratory muscles. The cough happens, with its typical sound, when the sudden opening of the glottis occurs with transient and fast expiratory airflow. A typical cough sound signal consists of three phases as shown in Fig. \ref{coughstructure}:
\begin{enumerate}
    \item a rapid explosive phase which is characterized by an initial burst of emerging sound yielding  a high frequency due to the vibrations produced by the forced air flux in the airway and the bronchial narrowing places.
    \item an intermediate decaying phase with a steady-state flow while the the glottis is wide open. This phase gives the duration to the whole cough. The presence of sputum may add a higher frequency component to the sound.
    \item a voiced phase (not necessarily present in all coughs) is characterized by a narrowing of the glottis again leading the vocal cords to get close to each other.
\end{enumerate}
The first two phases are ubiquitous across all coughs and will be useful in determining the start and end of a cough using an energy-based criteria. The majority of the coughs duration is around 400ms (Here we chose 50 ms) \cite{7570164}. For instance a dry cough is characterized by the absence of any mucus or sputum \cite{murata1998discrimination}. That is, all the three phases are visible in a
dry cough signal. Initially, a burst of high energy is observed followed by less energy in the second phase at higher frequencies. 
However in the case of a wet cough sound signal more energy is observed in Phase 2 at higher frequencies. Typically, a wet cough, symptomatic of bronchitis, asthma, and pneumonia, is produced by inflammation and secretion of mucus and sputum in the lower airways caused by either a bacteria or a virus \cite{murata1998discrimination}.

\begin{figure*}[]
\begin{center}
    \includegraphics[width=.9\linewidth]{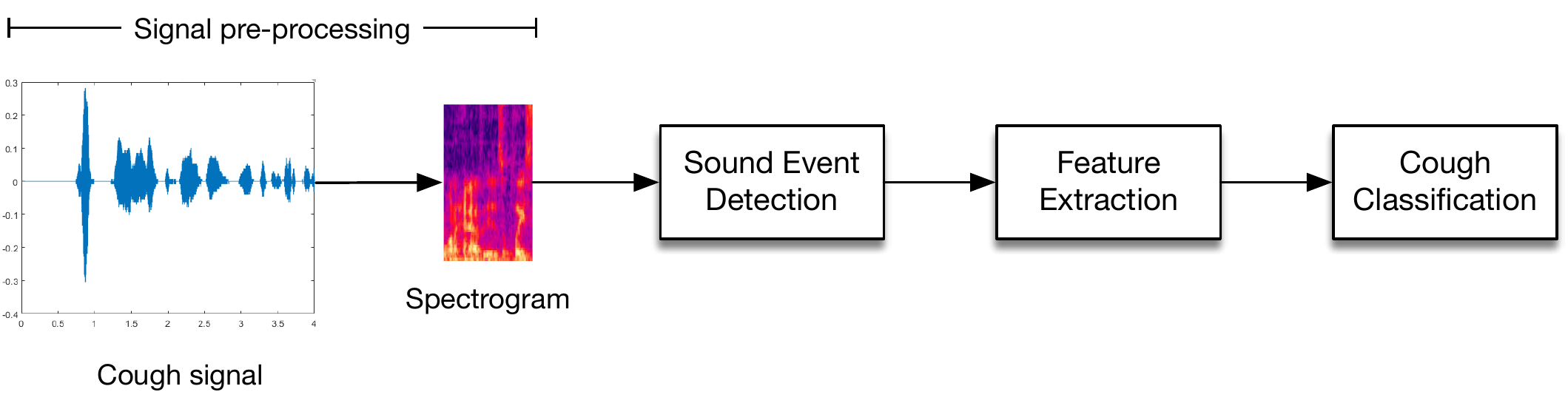}
    \caption{Cough audio processing.}
    \label{coughanalysis}
\end{center}
\end{figure*}

\subsubsection{Cough detection}
One of the first tasks of cough audio analysis is to be able to detect and identify a cough signal. Several research studies have addressed cough detection using different methods. For instance, in \cite{barry2006automatic}, the authors used Linear Predictive Coding and Mel-Frequency Cepstral Coefficients to model the sound of coughs, and used a Probabilistic Neural Network to classify time windows as containing or not a cough. Other researchers have used Hidden Markov Model (HMM) with MFCC to be fed as input \cite{matos2006detection}.

\begin{figure*}[]
\begin{center}
    \includegraphics[width=0.9\linewidth]{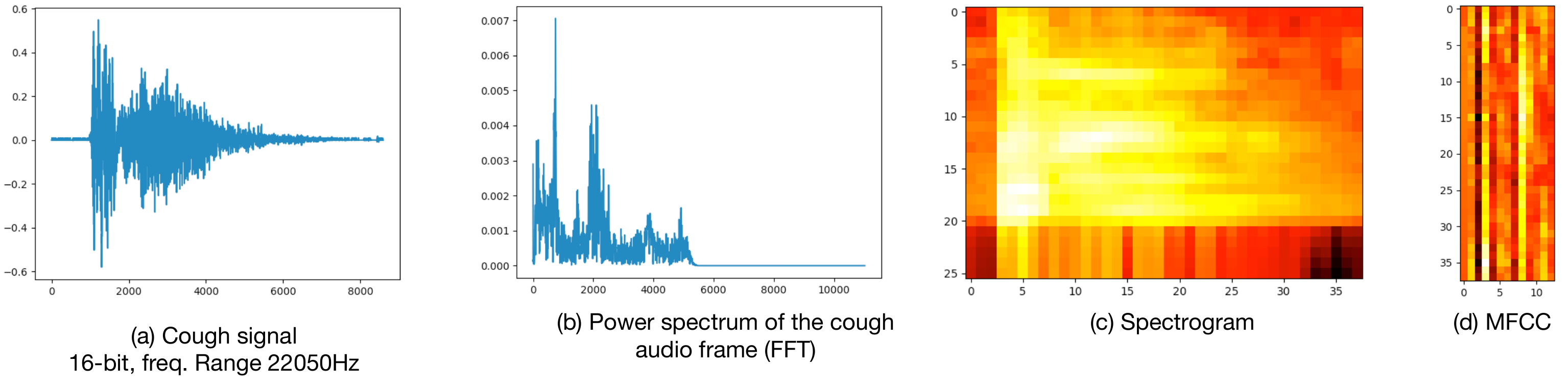}
    \caption{Cough time-frequency representation analyses.}
    \label{coughanalysis}
\end{center}
\end{figure*}

\subsubsection{Source separation}
This process consists of cleaning the cough sounds dataset by filtering out all the interferences and the environmental noise from the audio frames and keeping only the relevant frames to the cough. of audio separation consists of extracting the cough sounds. There are several methods for source separation including Independant ~Component Analysis (ICA) \cite{naik2011overview}, Blind Source Seprataion (BSS) \cite{naik2014blind} and Informed Source Separation (ISS) \cite{liutkus2013overview}.

\subsubsection{Feature selection and discriminant analysis}
Feature extraction is an important step towards classification of cough sounds. However, creating a classification model with from a dataset with high-dimensionality is time consuming and may converge to a local minima given the large search space. Therefore, selecting a reduced set of relevant features in an audio sample can improve immensely the performance generating a classification model \cite{he2013feature}. 
There are many techniques for feature selection including Shannon Entropy (SH), Fisher score, Mel-Frequency Cepstral Coefficients (MFCCs) and Zero Crossing Rate (ZCR). However, MFCC method has gained popularity due to its efficiency in the analysis of speech and sound signals in general, and is therefore opted for in the analysis of cough sounds. Features selection during the pre-processing of cough sounds is two-fold: first, reducing the dimentionality for the feature matrix classification, and second, extracting the most dominant information present in the cough sound. Prior to feature extraction, any noise must be filtered out of the cough sound. The quality and performance of the classification process depends strongly on how well the process of features extraction has been done. 

\subsubsection{Mel-frequency cepstral coefficients (MFCC)}
MFCCs are widely used in audio processing for sound pattern recognition. They are developed with the assumption that sounds are produced by glottal pulse passing through vocal tract filter. The MFCCs of a signal are a small set of features, usually about 10-20, which concisely describe the overall shape of a spectral envelope. MFCC analysis shows the power spectrum of the signal, and is often used to describe its timbre. MFCCs are calculated by taking cosine transform of a log-power spectrum on a non-linear Mel-scale of frequency. Cough sound signals are complex in nature, which indicates that the respiratory tract carries vital information and contributes the tract of cough signals from its substructure. A cough signal contains segments from each, Shannon entropy is calculated. Several techniques can be used for the discrimination between voiced and unvoiced speech signals. The signal is divided into small parts and calculated how many times signals are crossing zero axis.

The use of Mel scale is motivated by the fact that the features of an audio signal are better discerned by a human ear at low frequencies than they are at high frequencies. Therefore, Mel scale ensures that signal features match more closely what humans hear. The Mel scale is obtained using the following conversion formula:
\begin{equation}
    M(f)=1125\ln{(1+\frac{f}{700})}
\end{equation}
\begin{equation}
    M^{-1}(m)=700(\exp{\frac{m}{1125}})-1
\end{equation}
MFCCs  calculated following few steps starting by framing the signal into shorter frames (20-40 ms frames). That is, for a signal sampled at $f_s$ Hz, and a standard 25ms, the frame length is equal to $0.0025\times f_s$ samples. For each sample, the periodogram estimate of the power spectrum is calculated. The Mel filter bank is then applied to the power spectra and the sum of the energy in each filter is calculated. The next step consists of calculating the logarithm of all the  filter bank energies, and the discrete cosine transform (DCT) of the result. From the computed DCT, only coefficients 2 to 13 are retained as Mel coefficients. 
\vspace{-8mm}
\subsubsection{Cough Classification and Machine Learning}
Classification of cough sounds is a helpful tool for identifying the underlying cause of coughs. Several methods for automatic cough classification have been developed to identify various cough types and the thus the pulmonary disease. The most common of neural networks for deep learning include: Deep Neural Network (DNN), Convolutional Neural Network (CNN), Recurrent Neural Network (RNN), and Fuzzy Deep Neural Network FDNN).


\subsection{Proposed System Architecture}
We aim to design a reliable user-friendly AI based system for early detection of COVID-19 and other respiratory illnesses. This integrated hardware/software system will have two main components:

1) a novel hardware composed of several sensors (e.g., microphone, and thermal imaging tool, and a cough sound-recording device),

2) an AI software for cough classification and flu type recognition.

Figure 5 shows all necessary hardware components to collect health data from healthy and unhealthy participants. In the following paragraphs, we will explain each component in details.  

\begin{figure*}[]
\begin{center}
    \includegraphics[width=1\linewidth]{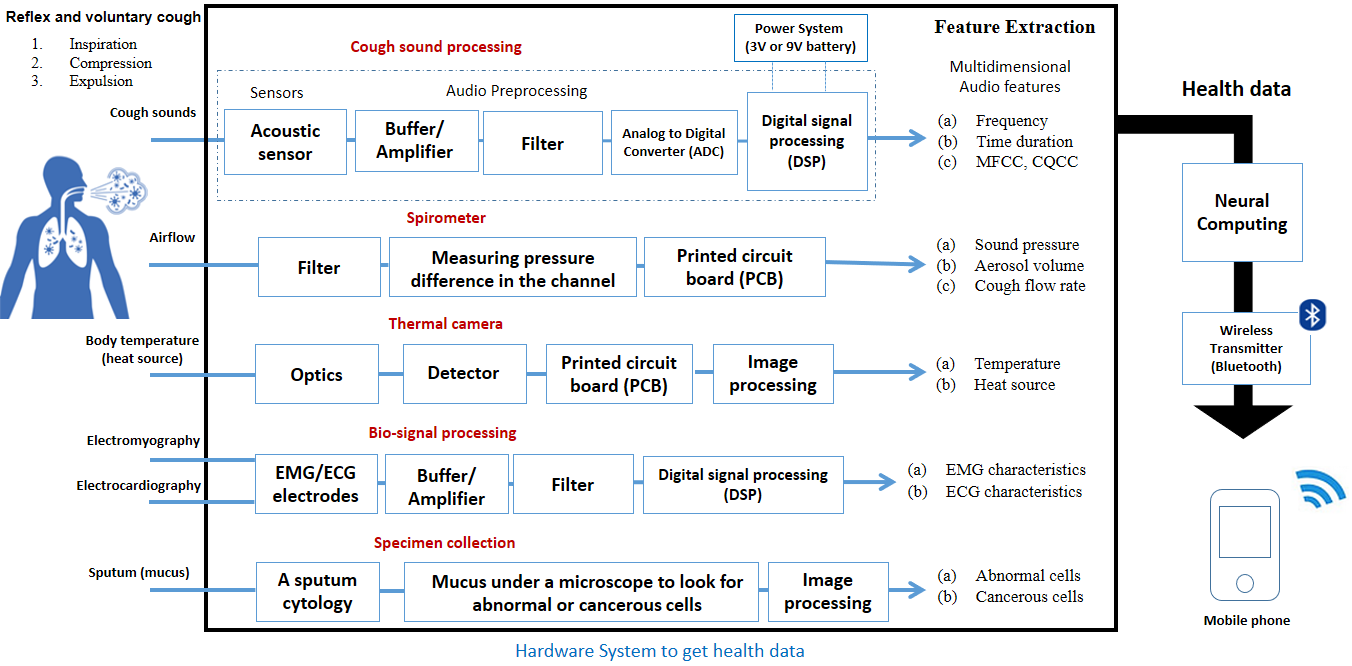}
\end{center}
    \caption{Hardware architecture of our proposed system to collect health data. The hardware contains relevant sensors to measure some parameters such as cough sound, breathing, body temperature and so on to be translated to relevant health data. Several sensors are needed to get relevant information from the human body (e.g., thermal camera, acoustic sensors, accelerometers, portable piezo contact microphones, airflow sensors, EEG/ECG electrodes). }
    \label{Hardware}

\end{figure*}

\subsubsection{Collecting data}
Collecting data strategy is one of critical phases to build a trusted diagnosis model. The data can be collected using specific sensors for each input. For example, collecting cough sound requires a simple microphone for recording the sound via smartphone app or a web browser. However, it is still a challenge to record big data for each respiratory illness. One way is to kindly ask people over the world to volunteer their cough sounds via online platforms. We may also collect basic demographics, medical history, and a few seconds of reflex and voluntary coughing samples. The participants should ensure that recording the sounds was in a quiet environment to avoid noisy sounds. The participant's anonymity and privacy are protected and no personal information is collected. If agreed on by the participant, the location may be  collected as location information is particularly useful to draw a map of respiratory illnesses. The collected data may be made available as open source to be used for promoting science and fighting respiratory diseases. Therefore, a process of ethical approval and clear informed consent from the participants may be required before before and during the collection.  
For developing an AI diagnosis for respiratory illnesses, many information can be recorded as inputs such as cough sound, body temperature, airflow, EMG, ECG, and mucus.    

\subsubsection{Cough sound recording}
The proposed system architecture is designed in order to extract useful clinical information from many inputs such as cough sounds. For recording audio of people coughing, breathing (which indicate labored or irregular breathing) or even talking, we need a microphone to convert acoustical energy (sound waves) into electrical energy (the audio signal). Then, we need amplifier, analog to digital converter (ADC), and digital signal processing (DSP) for prepossessing phase.

Cough characteristics and its acoustic features depend on the velocity of airflow, dimensions of the vocal tract and airways, and location of sound generated. Amplitude of the cough sound, intensity, duration, frequency, time-frequency representation (spectrogram which is a visual representation of the spectrum of frequencies of a signal as it varies with time.), the mel frequency cepstral coefficients (MFCC), the constant-Q cepstral coefficients (CQCC), and so on can be used for feature extraction phase as cough sound pattern. However, if the cough sound was not recorded in quiet room then blind source separation and independent component analysis can be used to find the right signal. 
\vspace{-4mm}
\subsubsection{Spirometer}

A spirometer is an apparatus for measuring the volume of air inspired and expired by the lungs of healthy or unhealthy participants. It measures ventilation, the movement of airflow into and out of the lungs which a high indicator for early diagnosis of any respiratory illnesses. For this subsystem, the input is airflow and the outputs could sound pressure, aerosol volume, and cough flow rate. 

\subsubsection{Thermal camera}
The infrared thermal imaging camera can be used for detecting elevated body temperatures which may indicate the presence of a fever, a symptom of many respiratory illnesses such as COVID-19. Detecting people with a potential fever may contain or limit the spread of many contagious respiratory illnesses through identification of infected individuals showing fever symptoms. Thermal cameras are passive devices that don’t emit any radiation, rather they use infrared radiation emitted form objects (human body in this case) to deliver high-resolution images without the need of any additional illumination. These cameras provide a visual map of skin temperatures in real time. They allow the operator of a critical public infrastructure (e.g., airports, train stations, and schools) to no-invasively scan the crowd or individuals to avoid the spread of some infectious disease. 

\subsubsection{EMG/ECG processing}

COVID-19 and many respiratory illnesses may cause trouble breathing, liver problems or damage, heart problems, and kidney damage. In most cases, the lungs might become inflamed, making it tough for patients to breathe. This can lead to pneumonia, an infection of the tiny air sacs (called alveoli) inside the lungs where the blood exchanges oxygen and carbon dioxide. However, during respiratory biofeedback, we can place some sensors or electrodes around the abdomen and chest to monitor the breathing patterns and respiration rate. 
Electrocardiogram (ECG) can be used for measuring the heart rate and how your heart rate varies (e.i., atrial and ventricular depolarization and repolarization are represented on the ECG as a series of waves PQRST: the P-wave, the QRS complex, and the T wave). In addition, electromyogram (EMG) can be also used for monitoring the electrical activity that causes muscle contraction around heart or chest wall movements or even for getting cough sound from the throat.

\subsubsection{Sputum cytology}
A sputum cytology is used for testing lung secretions or phlegm to look if there are some cancerous cells. The patient coughs up a sample of sputum (mucus), which is checked under the microscope to identify possible cancer cells or determine whether abnormal cells are present. However, using automated sputum cytometry such as LungSign test for lung cancer, lung ultrasound for pneumothorax, or optical automation for sputum cytology can be also useful compared to the conventional cytology.

\subsubsection{Artificial intelligence based algorithm}
After extraction some relevant patterns from system inputs and building a feature vector or matrix, we will give the data to some machine learning or deep learning methods (e.g., support vector machine (SVM), artificial neural network (ANN), convolutional neural network (CNN)). However, machine learning algorithms almost always require structured data, while deep learning networks rely on layers of ANN. 
Figure 7 shows software design of the proposed diagnosis system from collecting data until sending the health data with the preliminary diagnosis to the physicians. 
This software design has two phases: offline and online. For offline analysis, all collected data will be used for building a classification model by dividing the data into 80-90\% training data and 20-10\% test data. For classification method, deep learning is the best choice for big raw data. However, the number of class of respiratory illnesses is very high. Therefore, fuzzy logic might be an option to give a percentage of truth of each class. The model will be made based on many participants' data to be able to generalize it later in the testing phase. In addition, normalization of the data is also important. After fixing all parameters of the supervised classifier and building the model, we will test it on new data to check its performance. 
For real-time analysis, an online classification algorithm based on offline model (or dynamic model) is needed. The classification result of any respiratory illnesses will be display for the user on the smartphone and it can be sent to experts via cloud service. Further details on the implementation of different machine learning algorithms for cough classification is presented in the subsequent sections.
\renewcommand{\theenumi}{\alph{enumi}}
\begin{enumerate}
\item Deep Neural Network

Deep neural networks are distinguished from the more common single-hidden-layer neural networks by their depth. A deep neural network is a feed-forward, artificial neural network that has more than one layer of hidden units between its inputs and its outputs. Each hidden unit typically uses the logistic function to map its total input from the layer below that it sends to the layer above. In these systems the audio input is typically represented by concatenating MFCCs computed from the raw waveform. MFCCs are better adapted to facilitate discrimination with Hidden Markov Model (HMM). HMMs are used here as a decoder to capture tamporal information of the audio signals. DNNs also requires a fine-tuning step, which is a back propogation tuning that will allow the prediction of the observation probability associated with each HMM states.
\\
\item Convolutional Neural Network

Convolutional neural networks, like in neural networks, learnable kernels (filters) receive the audio spectral  features as input, takes a weighted sum over them, pass it through an activation function and responds with an output. A CNN often consists of a series of convolutional layers interleaved with pooling layers, followed by one or more dense layers. Pooling layers added on top of these convolutional layers can be used to downsample the learned feature maps.  A fully convolutional network (FCN) is a variant of CNN extended with fully connected layers. For feature extraction, we use a logarithmic (log)-scaled mel-spectrogram with an appropriate number of components (bands) covering the audible frequency range of a cough (0-22050 Hz), using a window size and a hop size of the same duration based on the number of samples. As mentioned earlier in this paper, the conversion to log-scaled mel takes into account the fact that human ear hears sound on log-scale, and closely scaled frequency are not well distinguished by the human Cochlea. 

\begin{figure*}[]
\begin{center}
    \includegraphics[width=0.9\linewidth]{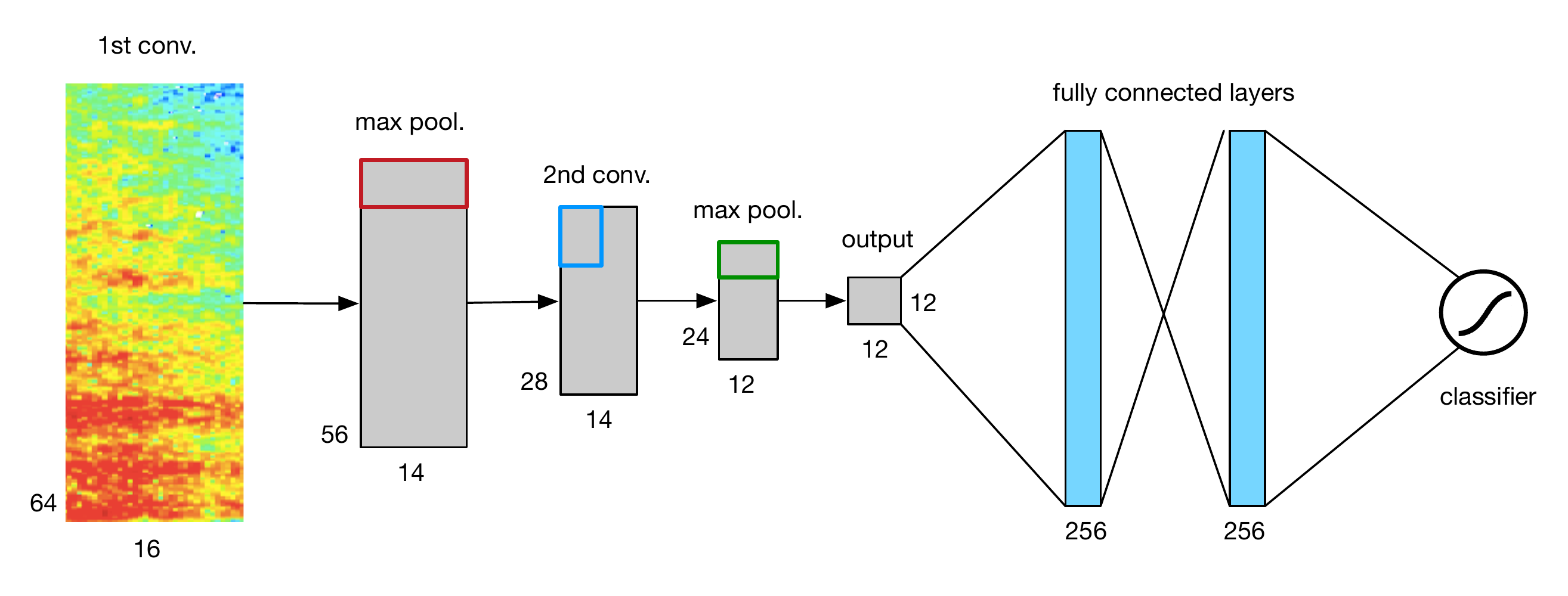}
    \caption{Example of a CNN architecture for a cough sound classifier.}
    \label{cnn}
\end{center}
\end{figure*}

     \item Recurrent Neural Network
     
RNNs allow cyclical connections in a feed-forward neural networks, which allows them to incorporate contextual information from previous input, and remember past inputs in the network’s internal state. This property makes RNNs an attractive choice for sequence to sequence learning. Compared to CNNs, RNNs follow a different approach for modeling sequences: They compute the output for a time step from both the input at that step and their hidden state at the previous step. This inherently models the temporal dependency in the inputs, and allows the receptive field to extend indefinitely into the past. For offline applications, bidirectional RNNs employ a second recurrence in reverse order, extending the receptive field into the future. Long short-term memory networks (LSTMs) are a variant of RNNs that exploit the contextual information over longer time intervals to map the input sequence to the output. LSTMs known to be efficient at learning temporal dependencies, and they are applied in a variety of areas such as , such speech recognition and synthesis. They are also used with some success when combined with CNN as front-end (CRNN), for video classification. Although, the applicability of LSTM for sound classification has not been fully investigated, they can be very beneficial given the temporal properties embedded in a cough sound.
\\

     \item Fuzzy Deep Neural Network
     
Integrating fuzzy inference systems into deep learning networks is one of hybrid classification approaches that have proven to be effective methods for making highly accurate predictions from complex data sources in the fuzzy logic domain (i.e., fuzzy sets, fuzzy rules). This combination aims to model vague notions with rigorous mathematical tools and rejects the principle of bivalence for pattern recognition, classification, regression or density estimation. Based on this aspect, many methods can be produced such as combining convolutional neural network with fuzzy logic which called the Fuzzy Convolutional Neural Network (FCNN). 
Both neural networks and fuzzy systems have no mathematical model necessary. Neural networks are able to learn from scratch using several learning algorithms but fuzzy systems are based on apriori knowledge and not capable to learn. Neural networks based on black-box behavior but fuzzy systems are based on simple interpretation and implementation. Therefore, combining both approaches aims to unite advantages and exclude disadvantages.

\begin{figure*}[]
\begin{center}
    \includegraphics[width=0.9\linewidth]{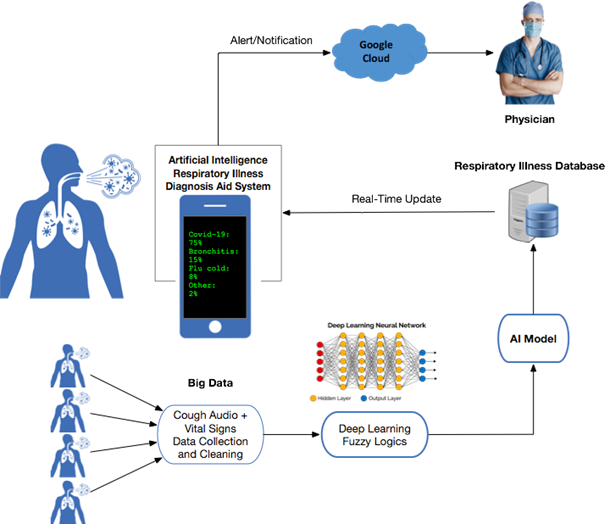}
\end{center}
    \caption{Software design of the proposed diagnosis system. This system includes all phases with the possibility of sharing the collected stored database and the preliminary diagnosis results with local and international researchers/medical doctors in the final phase of the product. The sharing approach will lead to develop more robust open-source algorithms to be used for detecting and classifying respiratory illnesses.}
    \label{Software}

\end{figure*}

\end{enumerate}

\subsection{Cloud hosting solutions}

Cloud is important for storing collected raw data and/or useful information for improving the decision-making of the physicians. The main benefit of this approach is to extend the use of
our proposed system to facilities that people with respiratory illnesses to send their health data and early diagnosis to medical experts with high confidentiality.

\section{Conclusions}
\label{Conclusions}

The proposed architectural design is composed of two major components. The first one consists of devising a process of collecting the data related to the vital signs of the patient and the recorded cough sound, breathing and so on. The second part consists of building an integrated hardware/software system for end-users intended to capture a patient’s vital signs along with the sound of the coughing. The hardware part of this system includes all the sensing devices whereas the software part contains the data processing as well as the prediction algorithms, which predicts the type of the respiratory illness including COVID-19. The prediction algorithm relies on the sate of the art AI model available in the respiratory illness database in the cloud. 
For the proposed system to gain traction in the market and be adopted by the public and healthcare systems, a number of KPIs need to be achieved and maintained, including efficient use of our proposed system, effectiveness and timeliness of early diagnosis, diagnosis costs, patient wait time, data transparency, decision-making errors of the used classifier, and patient safety and satisfaction. However, identifying the right KPIs, track them, and organize them in a logical, coherent and useful way, is critical step to achieve a good system evaluation, especially for developing medical devices. To evaluate this kind of systems, we shall conduct a pilot study and/or recruit volunteers (beta testers) to test the usability of the device and its software. Many other quality aspects such as reliability and performance efficiency can be assessed in the of evaluation process, where the involvement of medical experts will form a crucial role in this process.

The effective and efficient use of our proposed system is based on the collection of a combination of relevant health data for creating a feature vector to differentiate between multi-class respiratory illnesses. A timely diagnosis is a critical criteria for evaluating such a system; using advanced deep learning methods and dimensionality reduction lead for big data analysis will minimize the classification and diagnosis time. The proposed system is 
designed to be a portable and low cost diagnostic tool, supported with AI. Many countries in the world have limited resources, where this lab-free device will be handy. In addition, this system will increase data transparency between the patients and the physicians, and the ability to share data and diagnosis by the proposed system, with the medical experts via the cloud would improve timely intervention and reduce any false-positives of false-negatives.

\bibliographystyle{apsrev4-1.bst}
\bibliography{Paper}

\end{document}